\documentclass{PoS}

\usepackage{amsmath}
\usepackage{amsfonts}
\usepackage{amssymb}
\usepackage{dsfont}
\usepackage{physics}
\usepackage{braket}

\title{Towards the spectrum of the SU(2) adjoint Higgs model}

\ShortTitle{Towards the spectrum of the SU(2) adjoint Higgs model}

\author{\speaker{Vincenzo Afferrante}\thanks{Supported by the FWF doctoral school W1203-N16.}\\
        E-mail: \email{vincenzo.afferrante@uni-graz.at}}  

\author{Axel Maas\\
        E-mail: \email{axel.maas@uni-graz.at}}
        
\author{Pascal T\"orek\\
        E-mail: \email{pascal.toerek@uni-graz.at}  \\  
        
        Institute of Physics, NAWI Graz, University of Graz, Universit\"atsplatz 5, 8010 Graz, Austria}  

\abstract{Scalar particles in the adjoint representation of a non-Abelian gauge theory play an important role in many scenarios beyond the standard model, especially of GUT type. For such theories manifestly gauge-invariant, massless, composite vector particles have been predicted, even at weak coupling, using the Fr\"ohlich-Morchio-Strocchi mechanism. We use lattice gauge theory to investigate the simplest such theories, a single adjoint scalar coupled to an SU(2) Yang-Mills theory. The results support the existence of such a particle, in accordance with the prediction.
}

\FullConference{An Alpine LHC Physics Summit (ALPS2019)\\
		22-27 April, 2019\\
		Obergurgl, Austria}

\begin{document}

\section{Introduction}

In grand-unified theories (GUTs) the Abelian U(1) sector emerges as part of a non-Abelian gauge theory. However, a manifestly gauge-invariant formulation \cite{Frohlich:1980gj,Frohlich:1981yi,Maas:2017wzi} of particle physics holding beyond perturbation theory requires also in such a scenario that all low-energy degrees of freedom are gauge-invariant with respect to the full GUT gauge group \cite{Maas:2015gma}. Hence, QED needs to emerge as a low-energy effective theory of the composite states of the GUT. In particular, this requires the emergence of a uncharged, massless vector state, the low-energy photon, which is composite with respect to the GUT degrees of freedom. This appears to be the only way to conserve the perturbative GUT idea into one which takes the underlying subtleties of the quantum gauge theory into account \cite{Maas:2017xzh}.

Indeed, already in the standard model gauge-invariant composite states are on this grounds required to be the effective electroweak degrees of freedom \cite{Frohlich:1980gj,Frohlich:1981yi,Maas:2017wzi}. This works amazingly well, and can be explained by virtue of the Fr\"ohlich-Morchio-Strocchi (FMS) mechanism \cite{Frohlich:1980gj,Frohlich:1981yi}, which will be rehearsed in section \ref{s:fms}. Applying this mechanism to non-Abelian gauge theories with scalars in the adjoint representation indeed predicts the existence of such massless vector particle \cite{Maas:2017xzh}. This is already true for the simplest example, an SU(2) Yang-Mills theory with a single adjoint scalar.

Exploratory lattice simulations found hints of such state on tiny lattices \cite{Lee:1985yi}. We here substantially expand on these, and find evidence for the existence of such a state. In addition, we also confirm the analytical prediction for this state by explicitly testing the FMS mechanism.

\section{Continuum SU(2) adjoint Higgs theory}\label{s:fms}

	 The Lagrangian of an SU(2) Yang-Mills gauge theory coupled with a scalar field $\Phi$ in the adjoint representation, also known as Georgi-Glashow model \cite{Georgi:1974sy}, is
	 \begin{equation}
		\mathcal{L} = - \dfrac{1}{4}F^a_{\mu \nu} F^{a \mu \nu} + \tr [(D_\mu \Phi)^\dag(D^\mu \Phi)] +\mu^2 \tr \Phi^2 - \dfrac{\lambda}{2} (\tr \Phi^2)^2.
    \end{equation}
	The scalar field can be expanded as  $\Phi(x)  = \Phi^a(x) T^a$, where $T^a$ are the generators of the gauge group, and transforms under a gauge transformation as $\Phi(x) \rightarrow U(x) \Phi(x) U(x)^\dag$, $U(x)\in$ SU(2). The components $\Phi^a$ form a 3-dimensional real-valued vector.

    For our goals, we are interested in the Brout-Englert-Higgs (BEH) case. Then, in a suitable gauge, here minimal 't Hooft-Landau gauge \cite{Maas:2017wzi}, the scalar field can be split into a constant and a fluctuation part, i.e., 
			\begin{equation*}
			\Phi(x) = \braket{\Phi} + \phi(x) \equiv w\,\Phi_0 + \phi(x) \,.
			\end{equation*} Here,  $\Phi_0$ is the direction of the vacuum expectation value obeying $\Phi_0^a \Phi_0^a = 1$, and $w$ is its magnitude.  $\Phi_0$ can always be chosen inside the Cartan \cite{Bohm:2001yx}. 
			 The BEH effect generates a mass matrix for the gauge bosons
            \begin{equation*}
			(M^2_A)^{a b} = - 2 (gw)^2 \tr ([T^a,\Phi_0][T^b,\Phi_0]) \,.
			\end{equation*} 
			At tree-level  \cite{Bohm:2001yx}, for $SU(2)$ the only breaking pattern is $SU(2) \rightarrow U(1)$. Then, the U(1) subgroup gauge field is massless, and the masses of the SU(2) coset gauge bosons are $m^2_A= g^2 w^2$. In addition, one degree of freedom of the scalar field remains with mass $m_H^2 = \lambda w^2$. 
			
    However, the observable states need to be gauge-invariant \cite{Frohlich:1980gj,Frohlich:1981yi}, and are thus formed using composite operators. One suitable operator in the desired uncharged vector channel is \cite{Maas:2017xzh}
        \begin{equation}
		     O^\mu_{1^-} = \dfrac{\partial_\nu}{\partial^2} \tr [\Phi F^{\mu \nu}] \,\label{op}.
		\end{equation}
		To predict its mass, the FMS mechanism requires to expand it in both the vacuum expectation value and the couplings as \cite{Maas:2017xzh}
		\begin{equation}
		     O^\mu_{1^-} = - w \tr [\Phi_0 A_\perp^\mu](x) + \mathcal{O}(w^0) \,. 
		 \end{equation}
		 with $A_\perp^\mu = (\delta^\mu_\nu - \partial^\mu \partial_\nu/\partial^2  ) A^\nu $, the transverse part of the gauge field. Inserting this into its propagator yields for $\Phi_0^a = \delta_{a3}$ at tree-level
		\begin{align}
		    \braket{ O^\mu_{1^-}(x) O_{1^-,\mu}(y)} = \dfrac{w^2}{2}\braket{A^{3 \mu}_\perp(x) A^{3 }_{\perp,\mu}(y) }_{tl} + \mathcal{O}(w^0,g,\lambda) \,.\label{prediction}
		\end{align} 
		The subscript $tl$ stands for tree level. Comparing poles, this predicts a massless state in the channel, as required. This approach can be used to determine the whole spectrum \cite{Maas:2017xzh,Maas:2017wzi}.

\section{Lattice investigation}

We simulate the theory using standard lattice methods, including standard spectroscopical methods, which will be detailed elsewhere \cite{Afferrante:unpublished}. Here, we concentrate on the particularities of detecting a massless vector state. The lattice operator we use for the uncharged vector channel is \cite{Lee:1985yi}
\begin{equation}
        \label{eq:B_operator}
            O_{1^-}^i(x) = \dfrac{1}{ \sqrt{2\tr(\Phi^2) }} \Im\,\tr[\Phi(\vec{x},t)U_{jk}(\vec{x},t) ],
        \end{equation}
        where $U$ is the usual plaquette, and the indices $(ijk)$ are even permutations of $(123)$. This is the simplest discretization of (\ref{op}).
 
The correlator of an operator with definite momentum on an Euclidean lattice is \cite{Gattringer:2010zz} \begin{align}
	\big\langle O(\vec{p},t)O^\dag(\vec{p},t')\big\rangle 	&= A \cosh((t - t') E(\vec{p})) \left(1 + \mathcal{O}\left(e^{-a(t - t')\Delta E}\right)\right),
	\end{align}
	where $\Delta E$ is the splitting to the next channel. However, even in a finite volume, a massless vector particle cannot have a rest frame, as otherwise an artificial longitudinal degree of freedom would be necessary. Therefore, we need to switch to a moving frame. Thus, following \cite{Berg:1983is}, we give the operator a non-zero momentum via \begin{equation}
                B^j(\vec{p},t) =  \dfrac{1}{\sqrt{V_{\vec x}}} \Re\sum_{\vec x} O_{1^-}^j(\vec x,t) e^{i \vec{p} \cdot \vec{x}} \,.
            \end{equation}  The chosen momentum is the smallest one in the z direction on our symmetric lattices of size $N^4$
            \begin{equation}
                \vec{p}_z = \left(0,0,\dfrac{2 \pi}{N}\right) \,.
            \end{equation}  The correlator is then divided in a transverse and in a longitudinal component, using  \begin{align}
                C_\perp(t) &= \dfrac{1}{N}  \sum_{t'=0}^{N-1} \sum_{j=1}^2 \big\langle B^j(\vec{p}_z,t') B^{j\dag}(\vec{p}_z,t +t') \big\rangle \,,\\
                C_\|(t ) &= \dfrac{1}{N}  \sum_{t'=0}^{N-1}  \big\langle B^3(\vec{p}_z,t') B^{3\dag}(\vec{p}_z,t +t') \big\rangle \,.
            \end{align} 
            If the state is indeed entirely massless, the longitudinal part needs to vanish identically. Of course, if the operator has an overlap with massive states, this needs not to be the case.

\begin{figure}[tb!]
\begin{center}
    \includegraphics[width=0.5\textwidth]{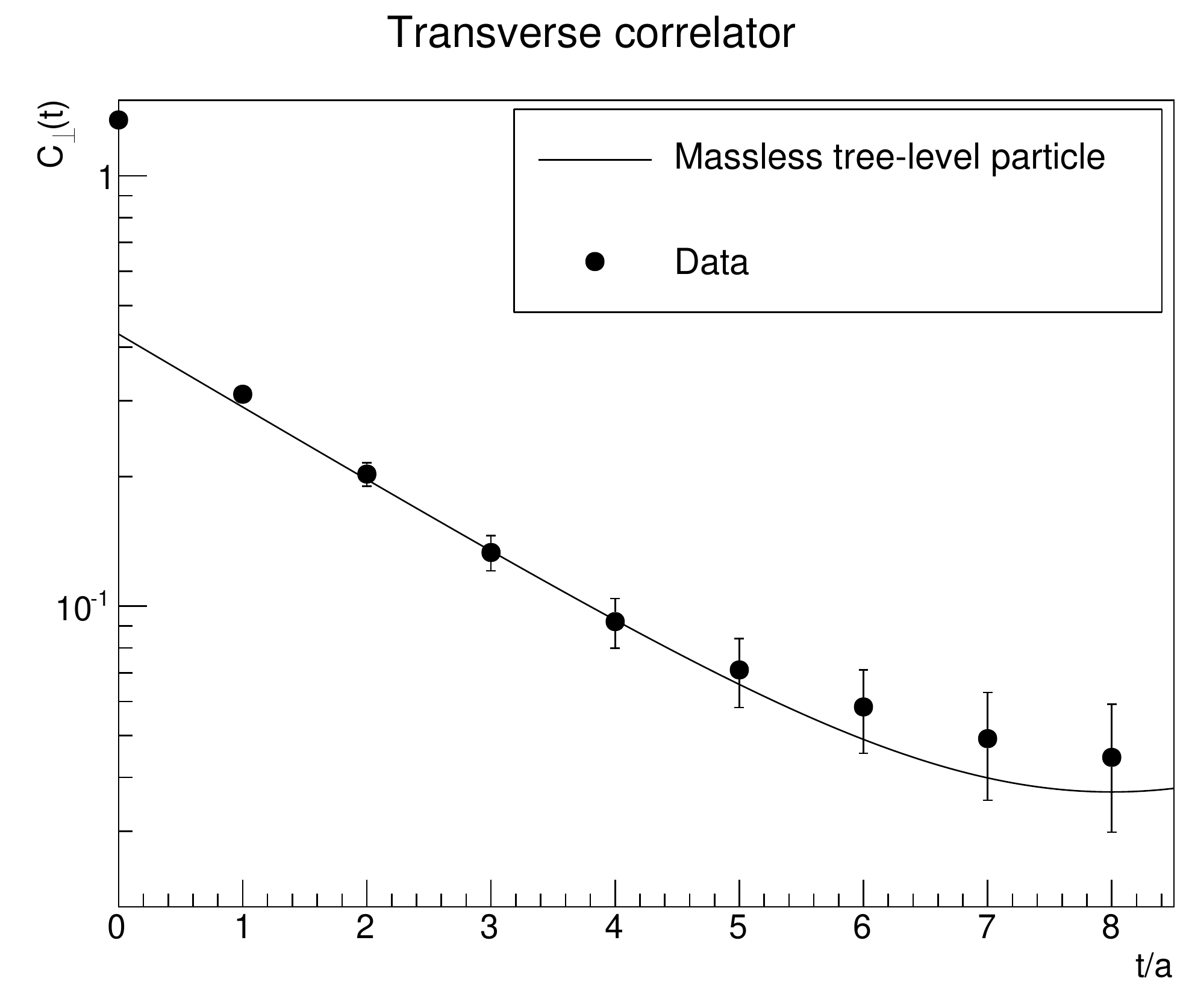}\includegraphics[width=0.5\textwidth]{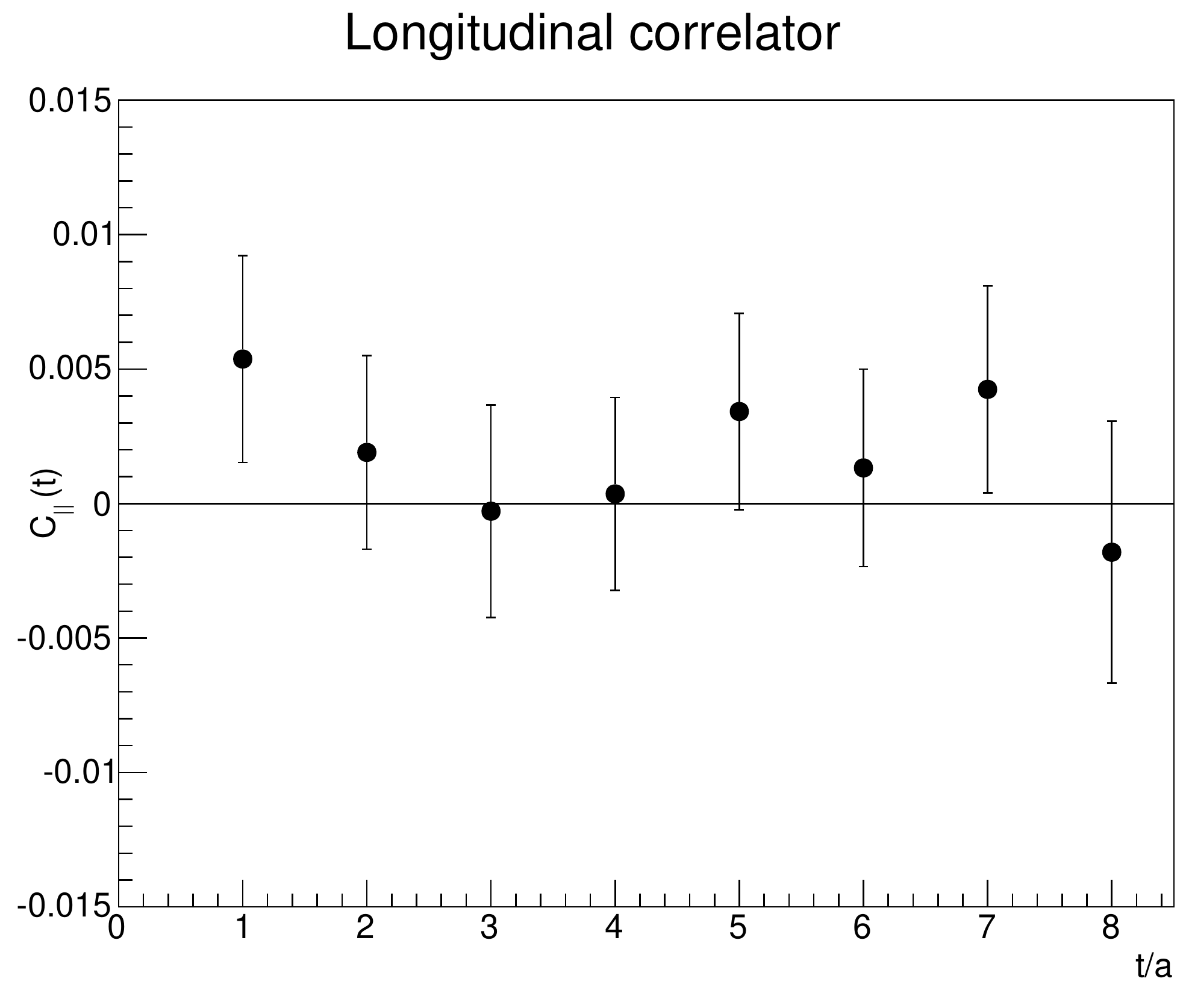}\\
    \includegraphics[width=0.5\textwidth]{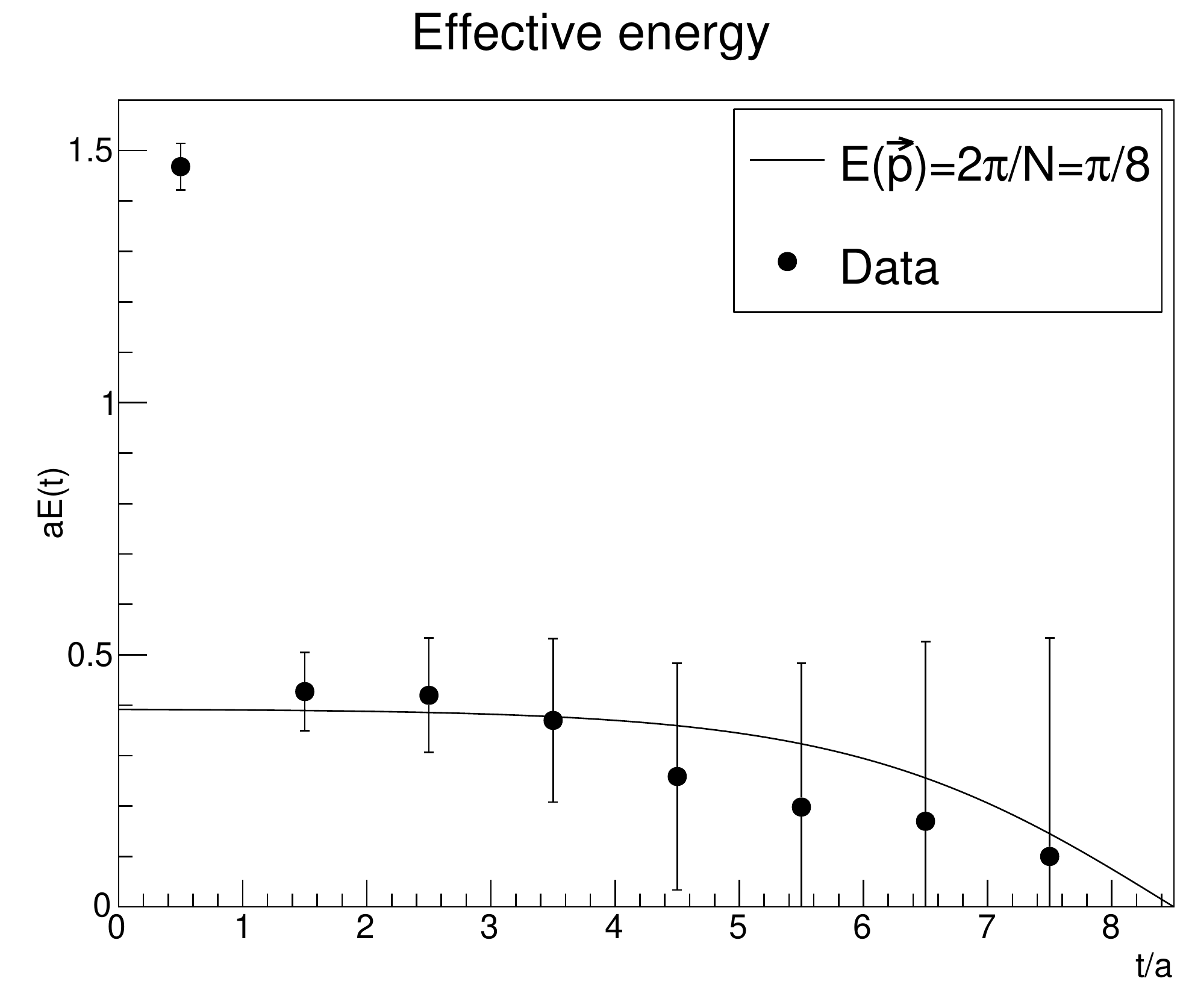}\includegraphics[width=0.5\textwidth]{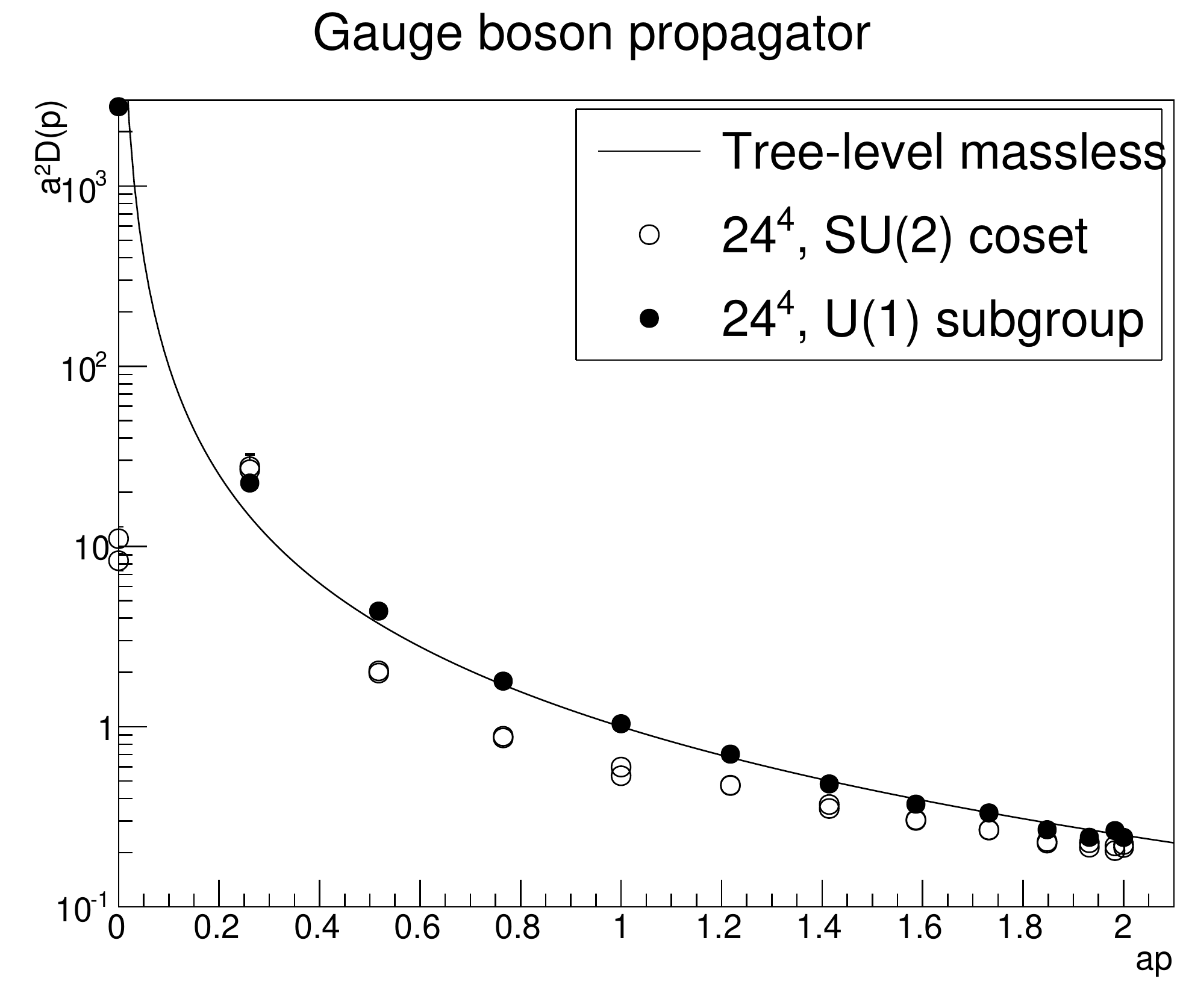}
    \caption{The transverse part (top-left panel) and longitudinal part (top-right panel) of the gauge-invariant vector correlator in a moving frame on a $N=16$ lattice. The corresponding effective energy is shown in the bottom-left panel. The gauge-dependent propagator, which requires less statistics, is shown in the bottom-right panel for $N=24$. Simulations parameters are $\beta=4$, $\kappa=3/4$, and $\lambda=1$ \cite{Afferrante:unpublished}.}
    \label{fig:B_parallel}
    \end{center}
\end{figure}

Preliminary results obtained for lattice parameters $\beta=4$, $\kappa=0.75$, $\lambda=1$, (see for a translation into continuum notation \cite{Afferrante:unpublished,Lee:1985yi}) are shown in figure \ref{fig:B_parallel} for two lattice sizes, $16^4$ and $24^4$. The absence of a signal in the longitudinal correlator is a strong hint of a massless state present in this theory. The effective energy $E(t)=-\ln(C_\perp(t)/C_\perp(t+1))$ obtained from the transverse correlator is also shown, together with the expected behavior for
$E(\vec p) = |\vec{p}_z|= 2 \pi/N$, showing again agreement. Hence, the results are in good agreement with the expectations for a massless particles.

We also show the gauge-dependent gauge-boson propagator in minimal 't Hooft-Landau gauge in figure \ref{fig:B_parallel}. The appearance of the massless Cartan mode is seen, while the coset modes show a typical \cite{Maas:2017wzi} massive behavior. This comparison of the left-hand-side and the right-hand-side of (\ref{prediction}) supports the FMS mechanism at leading order in this theory.

\section{Conclusions}

We have presented preliminary evidence that in a theory with an adjoint scalar a gauge-invariant, massless, composite vector particle exists, in agreement with the predictions of the FMS mechanism \cite{Maas:2017xzh}. This opens the way for an emergence of a gauge-invariant low-energy effective QED in a GUT setting. It also opens up interesting options for the generation of physical massless vector particles.

These results will require further detailed studies \cite{Afferrante:unpublished}, especially of the volume as a systematic lattice artifact. If established firmly, this will support the FMS mechanism as the suitable language to describe particle physics theories with BEH effect, and continue to pave the way for gauge-invariant model building.

\bibliographystyle{bibstyle}
\bibliography{bib}

\end{document}